# The Whittaker-Maxwell Scaffolding

**Scaffolding or template: what remains of a model when it is removed**

María Florentina Mendoza Urán
ORCID: 0009-0007-7882-6418

## Abstract

This paper discusses the metaphor, originating with Whittaker, according to which Maxwell's model of molecular vortices was the scaffolding by means of which the equations of electromagnetism were erected, removed afterwards without consequence for the building. It is argued here that the metaphor describes badly what happened, and it is shown that whoever upholds it today transcribes, within that same text, the quotations that belie it. The paper examines what was actually removed in 1865, which materials of the model still operate inside the equations under other names, and proposes in its place the figure of the template, taken from sheet-metal layout work. The discussion is situated with respect to the work of Daniel M. Siegel and Cameron Lazaroff-Puck, who maintain from the history of science the constitutive character of the model. It closes with a criterion for telling one case from the other.

## 1. Introduction

There is a sentence that has been deciding for more than a century how Maxwell is read.

Edmund Taylor Whittaker wrote it in 1910, thirty-one years after Maxwell's death, describing "A Dynamical Theory of the Electromagnetic Field", the work Maxwell had read before the Royal Society in December 1864 and which was published the following year: in it, he says, "the architecture of his system was displayed, stripped of the scaffolding by aid of which it had been first erected".

The scaffolding. The word caught on and settled in. And with it settled a complete way of telling the story: that Maxwell first built a mechanical model, a medium full of vortices with wheels between them; that this model served him to reach the equations; and that afterwards he removed it as one removes scaffolding once the building stands on its own. The equations remained; the rest was superfluous.

The image is convenient. It explains effortlessly why the equations still work today without anyone believing in vortices, and why they can be used without committing to any idea about what space is made of. It settles the matter at a stroke and leaves everyone at ease.

This paper argues that the image is false. Not exaggerated, not imprecise: false, in the concrete sense that it describes an operation which is not the one that took place.

Because scaffolding has one defining property, and that is that it does not touch the work. It is put up to reach where one cannot reach, it is taken down, and what has been built remains exactly as it would have been had it been done some other way. And from there comes the test, which is not whether the work stands without it, since a well-made work always stands, but whether what that something explained remains explained once it is removed. If it does, it was scaffolding. If not, it was something else.

What follows examines what was actually removed in 1865, what stayed inside the equations, and which figure correctly describes that relation. And it examines one further thing, not foreseen at the outset: what those who uphold the scaffolding reading say today, and whether what they say holds up against what they themselves transcribe.

## 2. The current reading

It is worth setting it out in its best version, and in its present-day version, not that of 1910.

In April 2015, on the occasion of the 350th anniversary of Philosophical Transactions, the Royal Society commissioned a series of commentaries on the most important papers published in its history. The one corresponding to Maxwell's 1865 work was signed by Malcolm Longair, of the Cavendish Laboratory, Cambridge.

Longair writes, in the abstract of the commentary, that the somewhat artificial mechanical models by which Maxwell had arrived at his field equations a few years earlier were stripped away.

And he develops the point further on: that Maxwell set about recasting the theory on a much more abstract basis, without any special assumptions about the nature of the medium through which electromagnetic phenomena are propagated.

There is the complete reading, and in the words of someone who knows exactly what he is talking about. Two concrete assertions: that the mechanical was stripped away, and that no special assumption about the medium remains.

Note further that this is not the rote repetition of an old phrase. Longair knows the specialist literature and cites it: in his commentary he expressly mentions the work of Daniel Siegel, who holds a different position and who will be examined further on. That is, the scaffolding reading is upheld today with full knowledge of the case, not through inertia.

So the question is not whether Whittaker was right a century ago. The question is whether those two assertions of Longair's withstand comparison with what Maxwell himself wrote in 1865, in "A Dynamical Theory of the Electromagnetic Field".

## 3. What Maxwell wrote

To answer that question there is no need to go looking for remote sources. It is enough to read the quotations that Longair himself reproduces in his commentary.

Summarizing Maxwell's approach, Longair writes that he therefore had to assume that there exists an aethereal medium pervading all bodies, modified only in degree by their presence; that the parts of that medium are capable of being set in motion by electric currents and magnets; and that this motion is communicated from one part of the medium to another by forces arising from the connections between those parts.

And he adds that central to his theory was the elasticity of the medium through which electromagnetic phenomena have to be propagated, which leads to the concept of the displacement current as a necessary part of the theoretical apparatus.

And he transcribes Maxwell saying that his theory may be called dynamical because it assumes that in that space there is matter in motion, by which the observed electromagnetic phenomena are produced.

So far: a medium, motion within it, elasticity, and the displacement current as a necessary consequence. But there is one more quotation, and it is the one that settles the matter. Longair reproduces it from paragraph 74 of Maxwell's paper:

> In speaking of the Energy of the field, however, I wish to be understood literally. All energy is the same as mechanical energy, whether it exists in the form of motion or in that of elasticity, or in any other form. The energy in electromagnetic phenomena is mechanical energy. The only question is, Where does it reside? On the old theories it resides in the electrified bodies, conducting circuits, and magnets, in the form of an unknown quality called potential energy, or the power of producing certain effects at a distance. On our theory it resides in the electromagnetic field, in the space surrounding the electrified and magnetic bodies, as well as in those bodies themselves, and is in two different forms, which may be described without hypothesis as magnetic polarization and electric polarization, or, according to a very probable hypothesis, as the motion and the strain of one and the same medium.

I wish to be understood literally. This is a man guarding against being read as though he were speaking in metaphor. And he writes it in 1865, in the very paper of which it is said that he had abandoned the mechanical.

## 4. The clash

Set the one beside the other.

Longair states that in 1865 Maxwell recast the theory without any special assumption about the nature of the medium through which electromagnetic phenomena are propagated.

And in the same commentary he transcribes Maxwell assuming an aethereal medium pervading all bodies, with parts capable of being set in motion and of communicating that motion to one another; he maintains that the elasticity of that medium was central to the theory; and he reproduces Maxwell himself saying that there is matter in motion in space, that electromagnetic energy is mechanical energy, that he wishes to be understood literally, and that it is very probably a matter of the motion and the strain of one and the same medium.

The two things cannot both be true. An aethereal medium, with motion, with central elasticity and with literal mechanical energy is an assumption about the nature of the medium. It is exactly that.

It is worth bringing in here the best of the opposing side, and it is Longair himself who supplies it a few lines earlier: he writes that Maxwell was well aware that the theory as propounded in "On Physical Lines of Force" was awkward, and that he regarded it as only a provisional and temporary hypothesis. These are expressions of Maxwell's own that Longair places in quotation marks, and this is the most serious objection that can be made to what is argued here.

But observe what the awkwardness applies to. Maxwell said it of the rolling contact between his cells, that is, of the machinery; and in that same passage, which Longair also records, he specifies that he does not bring it forward as a mode of connexion existing in Nature, and that even so it serves to bring out the actual mechanical connexions between known phenomena. Provisional that particular machine may well have been; real, the mechanical connexions that the machine served to bring to light. It is the same operation the letter to Tait will perform, and it is performed here by the very text cited to maintain the contrary.

It is worth saying what this does not mean. It does not mean that Longair is mistaken in his reading of the paper, which he knows far better than the present writer, nor that there is any intention to misrepresent. It means something more interesting and harder to see: that the inherited formula, that of the stripping away, has been applied over a body of material that belies it, and that the refutation lay in the very quotations chosen to illustrate it.

That is to say, the metaphor no longer describes the text. It replaces it.

And there remains, then, the question that really matters: if in 1865 there is still a medium, motion, elasticity and mechanical energy, what exactly was stripped away?

## 5. What was actually removed

The answer lies in Maxwell himself, and he gave it by letter.

On 23 December 1867, writing from Glenlair to Peter Guthrie Tait, and correcting an attribution in the draft of his Sketch of Thermodynamics, he compared his two works. The sentence is usually quoted clipped, with an ellipsis in the middle. It is worth reading it whole:

> There is a difference between a vortex theory ascribed to Maxwell at p 57, and a dynamical theory of Electromagnetics by the same author in Phil Trans 1865. The former is built up to show that the phenomena are such as can be explained by mechanism. The nature of this mechanism is to the true mechanism what an orrery is to the Solar System. The latter is built on Lagranges Dynamical Equation and is not wise about vortices.

What the ellipsis usually leaves out is the middle sentence, and it is the most important of the three. Maxwell is saying that his vortex machinery was not the true mechanism. He acknowledges it without evasion and with an exact comparison: that of a model which reproduces the phenomenon without being what actually happens.

But note what he says and what he does not. He does not say that there is no mechanism: he says that a true mechanism exists and that his was not it. And the comparison he chooses confirms this in the same movement. An orrery is neither a fiction nor an ornament: it is an apparatus built with the geometry of the orbits inside it, which reproduces their positions and their periods with accuracy sufficient to teach them and to predict them. No one would say it is superfluous for not being the sky. What Maxwell discards is the identity between his machinery and the real; not the existence of something real, nor the fact that his machinery reproduced it.

And in the third sentence he specifies what the 1865 work left out: it is not wise about vortices. The vortices, not the mechanism. That distinction has not gone unnoticed: Lazaroff-Puck has pointed it out, discussing the reading of Goldman and of Harman, who take that same sentence as proof of the purely mathematical character of the methodology of the 1865 work. And he adds something worth retaining: both admit that, despite the reform of his methods, the new conception of the electromagnetic field as a complicated mechanism remains explicitly grounded in a physical and mechanical analogy.

And it is exactly the distinction that the stripping-away formula erases. What was removed in 1865 was a particular piece of machinery: the vortex cells and the idle wheels between them. Not the medium. Not the motion. Not the elasticity. Not the mechanical character of the energy.

There is another letter usually cited to the contrary effect, and it is worth bringing it in whole. On 15 October 1864, writing to William Thomson, Maxwell says that he can now find the velocity of transmission of electromagnetic disturbances independently of any hypothesis, and that it is equal to v. Goldman and Harman take this as proof that by then he had freed himself from his hypotheses about the medium.

Three things are worth observing in it. The first is the context: this is not a programmatic letter, it is a laboratory letter. Maxwell proposes to Thomson an arrangement for weighing an electrostatic attraction against an electromagnetic repulsion, with aluminium discs, coils, a micrometer screw and a microscope; he talks to him about Cavendish's experiment, about the mechanical equivalent of heat and about the transparency of gold leaf. The sentence sits wedged between experimental arrangements.

The second is what the sentence says exactly. It speaks of finding a velocity, not of maintaining a theory; and that velocity he names v, which is the ratio between the electric and magnetic units, the same one that will reappear further on. He says that he can calculate that number without needing the hypothesis; he does not say that he has ceased to hold it. And its second half, which is almost never quoted, is a statement about the behavior of the medium: that the disturbances must be transverse to the direction of propagation or there is no propagation.

And the third is what stands a few lines earlier, in the same letter: Maxwell refers to the tendency in his rotatory theory of magnetism. In the possessive and in the present tense, less than two months before reading before the Royal Society the work of which it is said that he abandoned the mechanical. Harman's edition annotates that expression by referring back to the model of molecular vortices of "On Physical Lines of Force".

And there is one further piece of evidence, which turns out to be the most eloquent of all, because it lies in Longair's own commentary, and not through any oversight on his part: a few lines earlier he himself points to the mechanical origin of Maxwell's thought and to his technique of working by analogy. He transcribes another quotation from Maxwell of 1865, this one:

> the effect of the connexion between the current and the electromagnetic field surrounding it is to endow the current with a kind of momentum, just as the connexion between the driving-point of a machine and a fly-wheel endows the driving point with an additional momentum, which may be called the momentum of the fly-wheel reduced to the driving point.

A fly-wheel. A machine.

So that in the paper of which it is said that the mechanical models were stripped away there is an explicit mechanical analogy, and it is the commentator himself who reproduces it. He was not left without a machine: he changed machines. And that is no longer stripping away, but substituting.

## 6. Scaffolding or template

It remains to say which figure correctly describes what happened, since that of the scaffolding does not.

Return to the property that defines scaffolding: it does not touch the work. It is put up to reach where one cannot reach and it is taken down without leaving a trace. Think of a graffito painted from scaffolding. Remove the scaffolding and the graffito remains exactly the same. Not a mark of it is left, not a form, not a stroke. The work owes it nothing but access: it could have been painted from a ladder, from a cherry picker or hanging from a rope, and the result would be identical.

There is another figure that does describe the case, and it comes from sheet-metal layout work.

To make a sheet-metal bell one first draws a template: the development of the piece on the flat. Four trapezoidal faces, the height, the diagonals of the side lines, and the two- or three-centimeter tabs left for riveting or folding the corners. On that template the sheet is cut and it is bent where the template marks. When the bell is finished, the template is thrown away. The bell stands on its own and the template is nowhere to be found.

Why does it stand on its own? Because the right ingredients were used. And that is why the bell keeps the essence of the template: it is not an external aid that was taken away, it is what determined the result. Change the template and the piece changes. What is removed is the paper; what remains is the form that paper decided.

The difference between the two figures is not one of nuance. Scaffolding can be replaced by any other means of access without the work changing. A template cannot be replaced by another without a different piece coming out.

And the test of which of the two figures corresponds to each case is simple: look at what is left inside.

## 7. What stayed inside

It is worth seeing first what was on the table in 1861, because the ingredients of a work are better recognized if one knows where they came from.

The first thing there was, was a measured rotation. Faraday had discovered in September 1845, and published the following year, that a magnetic field rotates the plane of polarization of light. Thomson analyzed that result in 1856 and concluded that the cause of the magnetic action on light had to be a real rotation going on in the magnetic field. The vortices were not a happy notion that afterwards turned out to fit: they were the mechanical answer to a rotation that had already been measured and that had to be explained.

On that basis Maxwell draws his template in "On Physical Lines of Force", published in instalments in 1861 and 1862. And his materials are these: a sea of elastic vortex cells, contiguous, filling space, with their axes aligned along the lines of force; layers of small particles between them, transferring the spin from one vortex to the next without their jamming; a density of that sea; and a transverse elasticity, because the cells yield.

With those materials he reads his model. The magnetic field is not something that passes through the lattice: it is the spin of the lattice. The quantity H represents the vorticity, the pure spin. The quantity B is that same vorticity weighted by the density of the sea, and hence the relation between the two. Magnetic permeability is, in his model, the measure of the density of the sea of vortices. The electric current is the displacement of the particles at the junctions. And when those particles push against an elastic cell, they deform it a little before there is a current properly speaking: there is the displacement current.

Lastly, if the medium has density and has elasticity, a disturbance propagates through it as sound does through air. Maxwell applied the formula for the propagation velocity of a wave in an elastic medium with the density and the transverse elasticity of his sea, using the ratio between the electric and magnetic units measured by Weber and Kohlrausch, and out came almost exactly the velocity of light, obtained by a completely different experimental route.

Now remove the lattice, as it was removed. The equations stay standing and work just the same. But observe what is inside them.

H remains, which was the spin of the cells, and which today is the magnetic field and no more, without its being said what is spinning. The permeability remains, which was the density of the sea of vortices, and which today is a constant of the vacuum that is measured and tabulated, without its being said the density of what. The relation between B and H remains, which was vorticity times the density of the medium, and which today is a relation between two fields without a medium. The displacement current remains, which was the motion of the particles against an elastic cell before there was any current, and which today is the term needed for charge to be conserved. And the form of the velocity remains, which was that of a wave in a medium with density and elasticity.

And the spin remains, in the exact place where it was. In vacuum, of the four equations, two are divergence equations and two are curl equations. The curl is spin: it measures the circulation of a field around a point. And it does not appear just anywhere. The two divergence equations each speak of a single field. The two curl equations are precisely those that couple the two fields: the spin of one is tied to the variation of the other. The spin appears exactly where the fields couple, and it never appears where there is a single field alone.

It is all still there. What has disappeared is what it was made of. And that is why the equations work without the lattice: because they never described the lattice, they described what the lattice did. And what it did goes on happening, with lattice or without it.

## 8. Who had seen it before

Nothing said so far is an isolated discovery, and it is worth saying so plainly.

Daniel M. Siegel published in 1991, with Cambridge University Press, a study entitled "Innovation in Maxwell's Electromagnetic Theory: Molecular Vortices, Displacement Current, and Light", with a revised edition in 2003. His thesis, sustained through close analysis of the original texts, is that mechanical modelling played a crucial role in Maxwell's initial conceptualizations of the displacement current and of the electromagnetic character of light. He also maintains that Maxwell took the molecular vortex model very seriously, with ontological intent, and that although he later lost confidence in certain aspects of it and moved it to the periphery of his programme, he maintained his adherence to the central hypothesis. And he concludes that his two great innovations of that period arose from the theory of molecular vortices and reflected that context in their initial formulations.

Cameron Lazaroff-Puck published in 2015, in Archive for History of Exact Sciences, a work devoted precisely to the fly-wheel analogy in the 1865 paper. He begins by noting that this paper is usually remembered as the one that replaces the mechanical model with abstract mathematics, and that historians have considered the use of Lagrangian dynamics to be its sole important feature. Against that, he documents the often ignored mechanical analogy that Maxwell used to guide himself and his readers in the construction of his new equations: a weighted flywheel geared into two independently driven crank wheels, that is, a mechanical differential. And he shows how Maxwell used it to ground his study in clear mechanical conceptions and to structure the derivation of the equations. Longair's commentary is from the same year.

So the position exists in the literature, it is published by first-rank presses and journals, and it is prior. What does not yet exist is a figure to replace the old one, and that is why the scaffolding figure still occupies the language: one can demonstrate over and over that the model was essential and go on saying that the scaffolding was removed, because there is no other word to hand.

## 9. The criterion

From all the foregoing a general criterion follows, applicable beyond the case of Maxwell.

Faced with a model whose name has been removed, the question usually asked is whether it was needed. And that question is badly framed, because almost nothing is needed if all one wants is for the sums to come out. To a question of sufficiency one can always answer that it was not indispensable, and with that everything is said and nothing explained.

Nor does it help to ask whether the work stands without it, because a well-made work always stands. Nor whether the result could have been reached by another route, because almost always it could, and indeed sometimes it was: two years after Maxwell, and starting from a completely different conceptual basis, appealing to no medium and using retarded potentials, Ludvig Lorenz published an electrical theory of light that led to almost the same results (Kragh, 2018). That one can arrive at a place by another route says nothing about what stayed inside the first.

The correct question is another: what did that model answer, and does what it answered still have an answer after it is removed?

Apply it to the case. The vortex model answered why there are two magnetic quantities and not one, why a constant appears that measures a permeability, why the propagation velocity comes out of a density and an elasticity, and why the spin is where it is. Remove the model and observe what happens: the quantities remain, the constant remains, the form of the velocity remains, the spin remains. The answers do not.

And that is the criterion. If what the model explained remains explained without it, it was scaffolding: it gave access and nothing more, and it can be forgotten without loss. If on removing it its results are kept but its explanations are lost, it was a template: it determined the result, and its essence remains inside even though its name has vanished from the text.

The criterion matters because the operation recurs. Every time a theory abandons its substrate of origin and keeps its formalism, someone concludes that the substrate was dispensable. Sometimes it was. But it is worth distinguishing between what is removed because it is in the way and what is removed because it has already left its form in the work. Taking down scaffolding is not the same as throwing away a template.

# References

Faraday, Michael (1846). "I. Experimental researches in electricity.-Nineteenth series". Philosophical Transactions of the Royal Society of London, 136, 1-20. DOI: 10.1098/rstl.1846.0001.

Goldman, Martin (1983). The Demon in the Aether: The Story of James Clerk Maxwell. Paul Harris Publishing, Edinburgh. Pages cited through Lazaroff-Puck (2015).

Harman, Peter M. (1998). The Natural Philosophy of James Clerk Maxwell. Cambridge University Press, Cambridge. Pages cited through Lazaroff-Puck (2015).

Kragh, Helge (2018). "Ludvig Lorenz and His Non-Maxwellian Electrical Theory of Light". Physics in Perspective, 20, 221-253. DOI: 10.1007/s00016-018-0223-1.

Lazaroff-Puck, Cameron (2015). "Gearing up for Lagrangian dynamics: The flywheel analogy in Maxwell's 1865 paper on electrodynamics". Archive for History of Exact Sciences, 69(5), 455-490. DOI: 10.1007/s00407-015-0157-9.

Longair, Malcolm (2015). ""A paper... I hold to be great guns": a commentary on Maxwell (1865) "A dynamical theory of the electromagnetic field"". Philosophical Transactions of the Royal Society A, 373, 20140473. DOI: 10.1098/rsta.2014.0473.

Lorenz, Ludvig Valentin (1867). "On the Identity of the Vibrations of Light with Electrical Currents". Philosophical Magazine, fourth series, 34, 287-301.

Maxwell, James Clerk (1861-1862). "On Physical Lines of Force". Philosophical Magazine, fourth series, published in instalments: 21 (1861), 161-175, 281-291 and 338-348; and 23 (1862), 12-24 and 85-95.

Maxwell, James Clerk (1865). "A Dynamical Theory of the Electromagnetic Field". Philosophical Transactions of the Royal Society of London, 155, 459-512. Read before the Royal Society on 8 December 1864.

Maxwell, James Clerk. Letter to William Thomson, 15 October 1864. Document 235 in Harman, Peter M. (editor), The Scientific Letters and Papers of James Clerk Maxwell, volume 2. Cambridge University Press, 1995, pp. 176-181. From the original in the University Library, Glasgow.

Maxwell, James Clerk. Letter to Peter Guthrie Tait, 23 December 1867. Document 278 in Harman, Peter M. (editor), The Scientific Letters and Papers of James Clerk Maxwell, volume 2. Cambridge University Press, 1995, pp. 335-339. From the original in the University Library, Cambridge.

Siegel, Daniel M. (1991). Innovation in Maxwell's Electromagnetic Theory: Molecular Vortices, Displacement Current, and Light. Cambridge University Press, Cambridge. First edition, ISBN 0-521-35365-3. Revised paperback edition, 2003, ISBN 978-0-521-53329-4.

Thomson, William (1856). "Dynamical illustrations of the magnetic and the helicoidal rotatory effects of transparent bodies on polarized light". Proceedings of the Royal Society of London, 8, 150-158.

Weber, Wilhelm Eduard and Kohlrausch, Rudolf (1856). "Ueber die Elektricitätsmenge, welche bei galvanischen Strömen durch den Querschnitt der Kette fliesst". Annalen der Physik und Chemie, 99, 10-25.

Whittaker, Edmund Taylor (1910). A History of the Theories of Aether and Electricity. Longmans, Green and Co., London. First edition, in a single volume.

Whittaker, Edmund Taylor (1951). A History of the Theories of Aether and Electricity. Volume 1: The Classical Theories. Thomas Nelson and Sons, London, p. 255. Revised and enlarged edition, in two volumes; this is the one cited by Longair (2015).